\documentclass[twocolumn,superscriptaddress,aps]{revtex4}
\usepackage{amsmath}
\usepackage{amssymb}
\usepackage{graphicx}
\usepackage{color}
\usepackage[colorlinks=true,linkcolor=blue,citecolor=blue]{hyperref}
\usepackage{tabularx}
\usepackage{float}

\newcommand{\be}{\begin{equation}}
\newcommand{\ee}{\end{equation}}
\newcommand{\bea}{\begin{eqnarray}}
\newcommand{\eea}{\end{eqnarray}}
\newcommand{\bse}{\begin{subequations}}
\newcommand{\ese}{\end{subequations}}
\newcommand{\parent}{${\rm SrMnSb_2}$}
\newcommand{\Ksub}{${\rm (Sr_{0.97}K_{0.03})MnSb_2}$}

\newcommand{\Tn}{$T_{\textrm N}$}

\newcommand{\Cudop}{${\rm Sr(Mn_{0.9}Cu_{0.1})Sb_2}$}
\newcommand{\Zndop}{${\rm Sr(Mn_{0.9}Zn_{0.1})Sb_2}$}
\begin{document}

\author{Farhan Islam}
\author{ Renu Choudhary}
\affiliation{Ames Laboratory, Iowa State University, Ames, Iowa 50011, USA}
\author{Yong Liu}
\affiliation{Ames Laboratory, Iowa State University, Ames, Iowa 50011, USA}
\affiliation{Crystal Growth Facility, Institute of Physics, The {\'E}cole Polytechnique F{\'e}d{\'e}rale de Lausanne, CH-1015 Lausanne, Switzerland}
\author{Benjamin G. Ueland}
\author{Durga Paudyal}
\affiliation{Ames Laboratory, Iowa State University, Ames, Iowa 50011, USA}
\author{Thomas Heitmann}
\affiliation{The Missouri Research Reactor, University of Missouri, Columbia, Missouri 65211, USA}
\author{Robert J. McQueeney}
\author{David Vaknin}
\email{vaknin@ameslab.gov}
\affiliation{ Ames Laboratory, and Department of Physics and Astronomy, Iowa State University, Ames, Iowa 50011, USA}

\title{Controlling Magnetic Order, Magnetic Anisotropy, and Band Topology in Semimetals {\Cudop} and {\Zndop} }
\date{\today}

\begin{abstract}

Neutron diffraction and magnetic susceptibility studies show that orthorhombic single-crystals of topological semimetals {\Cudop} and  {\Zndop} undergo three dimensional  C-type antiferromagnetic (AFM) ordering of the Mn$^{2+}$ moments at $T_N = 200\pm10$ and $210\pm12$ K, respectively, significantly lower than that of the parent  SrMnSb$_2$ with $T_{\rm N}=297 \pm 3$ K. Magnetization versus applied magnetic field (perpendicular to MnSb planes)  below  $T_{\rm N}$ exhibits slightly modified de Haas van Alphen oscillations for the Zn-doped crystal as compared to that of the parent compound. By contrast, the Cu-doped system does not show de Haas van Alphen magnetic oscillations, suggesting that either Cu substitution for Mn changes the electronic structure of the parent compound substantially, or that the Cu sites are strong scatterers of carriers that significantly shorten their mean free path thus diminishing the oscillations. Density functional theory (DFT) calculations including spin-orbit coupling predict the C-type AFM state for the parent, Cu-, and Zn-doped systems and identify the $a$-axis (i.e., perpendicular to the Mn layer) as the easy magnetization direction in the parent and 12.5\% of Cu or Zn substitutions. In contrast, 25\% of Cu content changes the easy magnetization to the $b$-axis (i.e., within the Mn layer). We find that the incorporation of Cu and Zn in \parent\ tunes electronic bands near the Fermi level resulting in different band topology and semi-metallicity. The parent and Zn-doped systems have coexistence of electron and hole pockets with opened Dirac cone around the Y-point whereas the Cu-doped system has dominant hole pockets around the Fermi level with a distorted Dirac cone. The tunable electronic structure may point out possibilities of rationalizing the experimentally observed de Haas van Alphen magnetic oscillations.
\end{abstract}

\maketitle

\section{Introduction}
SrMnSb$_2$ belongs to a class of bulk topological semimetals ($A$Mn$Pn_2$;  $A =$ Sr, Ba, Ca; $Pn =$ Sb and Bi) where Sr-Sb layers are predicted to host Dirac fermions \cite{Park2011,Wang2011a,Wang2012,Wang2012a,Lee2013, Farhan2014}.  The degeneracy of the Dirac bands can be removed by breaking either time-reversal symmetry (TRS) or inversion symmetry  to produce Weyl fermions in the bulk of these systems. Time reversal symmetry can be broken by a magnetic field (internally or externally applied) which makes this class of materials intriguing as they possess exchange-coupled Mn$^{2+}$ layers, albeit, Mn-Mn coupling is primarily antiferromagnetic (AFM) and requiring net ferromagnetism (FM) to break the average TRS at the Sb site \cite{Liu2016}. It has been reported that Mn- and Sr- deficiencies in Sr$_{1-y}$Mn$_{1-z}$Sb$_2$ with $y \sim 0.08$ and $z \sim 0.02$ can generate a spontaneous spin canting that gives rise to weak-ferromagnetism \cite{Liu2017}.  As a result, Sr$_{1-y}$Mn$_{1-z}$Sb$_2$  is reportedly a Weyl semimetal that emerges by subjecting the Dirac fermions to a broken TRS via weak FM ordering \cite{Liu2017}.  Subsequent studies have reported much weaker or even absent FM in the stoichiometric SrMnSb$_2$ \cite{Ramankutty2018,Liu2019}.

\begin{figure}[b]
\includegraphics[width=3.3 in]{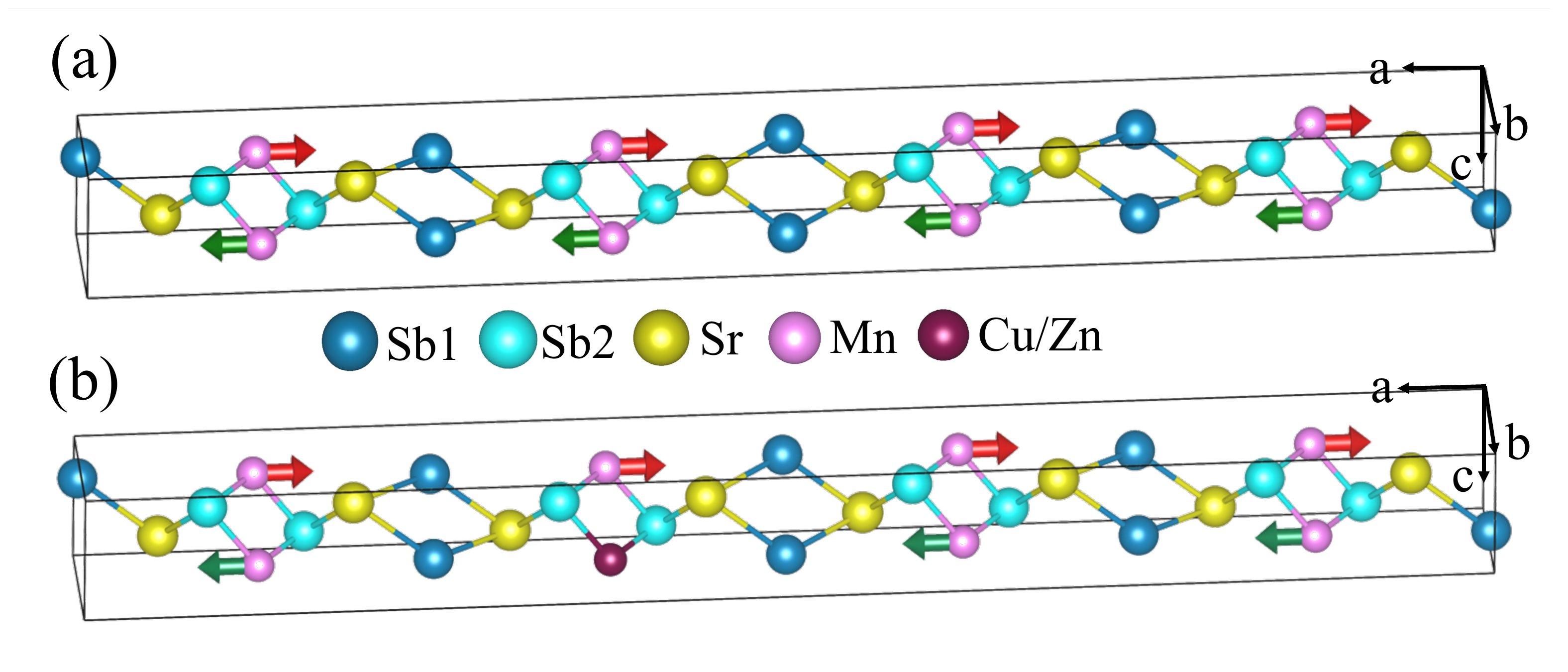}
\caption{Chemical and C-type AFM spin structure showing a $2\times1\times1$-supercell of (a) parent \parent\ and (b) 12.5\% Cu/Zn-doped \parent. Red and green arrows on Mn-atoms represent the spin direction of Mn moment along the $a$-axis. Note that this $Pnma$ orthorhombic structure has a long $a$-axis with planes of MnSb in the $bc$-basal plane. }
\label{Fig:structure}
\end{figure}

\begin{figure*}
\centering
\includegraphics[width=6 in]{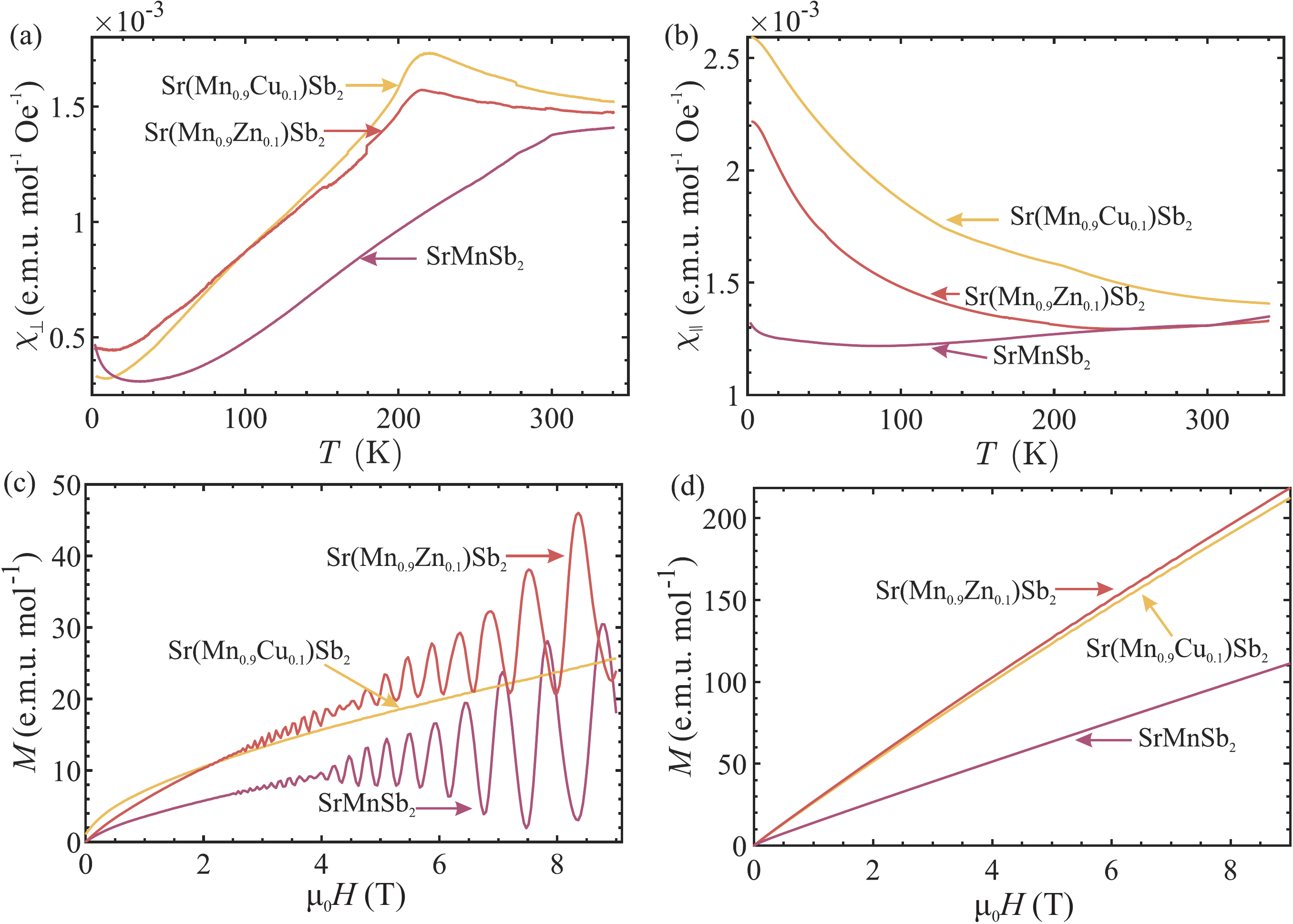}
\caption{Comparison of magnetic properties of parent compound \parent\ with doped  \Zndop\ and \Cudop. Temperature dependence of magnetic susceptibility  is shown with both (a) $H$ perpendicular to the MnSb planes  $\chi_{\perp}(T)$ and (b)  with $H$ in the MnSb planes $\chi_{||}(T)$. (c) Magnetization versus applied field perpendicular to the MnSb planes $M_\perp$  at $T=2$ K showing de Haas van Alphen (dHvA) oscillations for the parent and  \Zndop\ compounds but not for \Cudop. (d) Magnetization versus applied field in the MnSb planes $H_{||}$  at $T=2$ K exhibiting linear field dependence as expected from an AFM system with moments that are aligned perpendicular to the applied field.} 
\label{Fig:Mag}
\end{figure*}

Based on the interest to understand the role of defects, and the capability of carrier doping to tune topological semimetals, we recently reported neutron diffraction and magnetic susceptibility studies of hole-doped single crystal (Sr$_{0.97}$K$_{0.03}$)MnSb$_2$ confirming the three-dimensional (3D)  C-type type AFM ordering of the Mn$^{2+}$ moments at a slightly higher N{\'e}el temperature ($T_{\rm N}$) than that of the parent SrMnSb$_2$. These compounds form checkerboard-like AFM layers that are stacked ferro-magneticallly, as shown in Fig. \ref{Fig:structure} (a). de Haas van Alphen (dHvA) oscillations can provide a sensitive probe of the Fermi surface geometry and are extensively used to characterize topological semimetals. In {\Ksub}, magnetization versus applied magnetic field exhibits dHvA oscillations with a slight frequency shift as compared to the parent compound, presumably due to hole doping of the system \cite{Liu2019b}.  Here, we examine the effect of Cu and Zn substitutions of the Mn site - i.e., {\Cudop} and  {\Zndop} - by employing single-crystal neutron diffraction techniques and magnetic measurements. We find that the $T_{\rm N}$ is significantly lower in these systems compared to their parent compound. Moreover, whereas {\parent}, {\Ksub}, and {\Zndop} show dHvA oscillations, {\Cudop} does not.

We also report DFT calculations incorporating spin-orbit couplings to track the effects of Cu and Zn substitution of Mn on the band structure to rationalize their experimentally observed distinct behaviors. Previous band structure calculations combined with ARPES have shown that the vicinity of the Y-point of SrMnSb$_2$ has small Fermi pockets with linearly dispersing bands \cite{Ramankutty2018}. These pockets give rise to the dHvA oscillations in {\parent}. Therefore, we focus our DFT calculations on the effect that Cu or Zn substitutions for Mn have on the Y-point. Indeed, we find that Zn substitution does not significantly affect the Y-point of the parent compound, whereas Cu substitution does.

\begin{figure*}
\includegraphics[width=6 in]{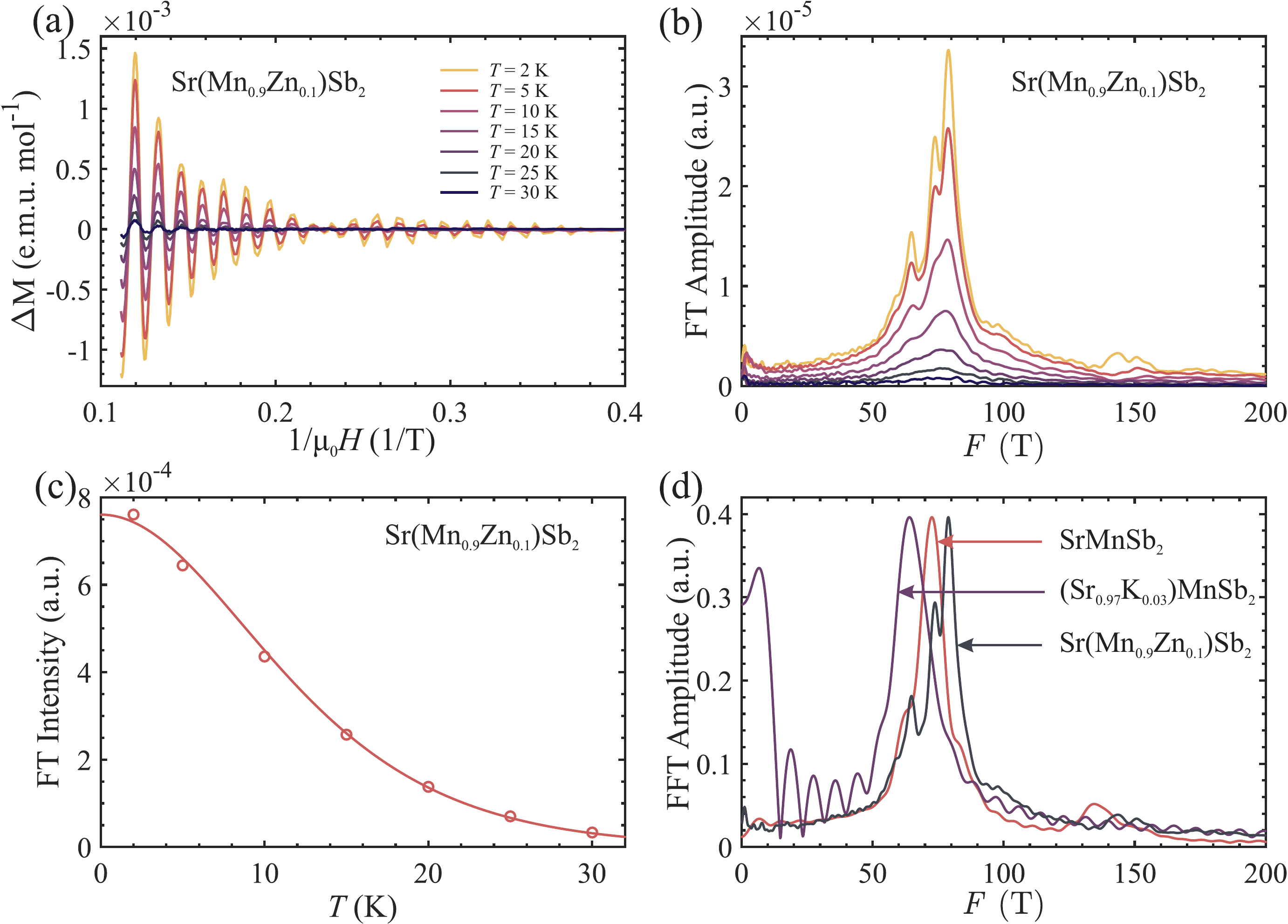}
\caption{(a) Corrected magnetization versus the inverse applied magnetic field ($1/H$) perpendicular to MnSb planes, $\Delta M(1/H)$, of \Zndop\ at various temperatures, as indicated, showing dHvA oscillations (data were corrected for an FM contribution as discussed in Refs.  \cite{Liu2019,Liu2019b}).  (b) Numerical FT spectra obtained from the $\Delta M(1/H)$ data shown in (a). (c) Temperature dependence of the integrated intensity (from 40 to 120 T) of the FT spectra shown in (b). (d) Comparison of  FT spectra at $T = 2$ K of the parent compound \parent, and the doped compounds \Zndop\ and \Ksub \cite{Liu2019b}.}
\label{Fig:dHvA}
\end{figure*}

\section{Experimental Details and Methods}
Plate-like single crystals of \Cudop\ and \Zndop\ were grown by a self-flux method. Cu and Zn substituted samples were prepared from a starting molar ratio of Sr:Mn:Cu(Zn):Sb=1:0.9:0.1:4.  The strontium dendritic pieces, manganese powder, copper powder, zinc pellets, and antimony chunks were weighed and loaded into alumina crucible in a glovebox. The alumina crucible was sealed in an evacuated quartz tube with backfilling 300 mbar Ar gas. The quartz ampoule was heated up to 1073 K. After a dwell of 6 hours, it was cooled down to 773 K at a cooling rate of 3 K/h. Then the furnace was switched off. The centimeter-sized plate-like single crystals were mechanically separated from the matrix. More details can be found in Ref. \cite{Liu2019}. The Cu and Zn content was determined by Energy-dispersive X-ray spectroscopy (EDS) measurements, which is consistent with the nominal doping.

X-ray diffraction (XRD) measurements of the ($h$00) reflections from the surfaces of the crystals were performed on a Bruker D8 Advanced Powder Diffractometer by using Cu K$\alpha$ radiation. These measurements indicate that the $a-$axis (corresponding to the crystallographic direction normal to the layers in the $Pnma$ structure) lies parallel to the surface normal of the plate-like crystals. 

We note that there are two inequivalent Sb atoms in the unit cell (labelled as Sb1 and Sb2 in Fig.\ \ref{Fig:structure}). The Sb1 atoms play a prominent role in the topological character of the system, contributing to the formation of linear Dirac bands near the Fermi level ($E_F$) at the Y-point \cite{Farhan2014,Liu2017}. The Sb2 atoms are bonded to the Mn to form corrugated MnSb layers, common to a plethora of layered Mn based pnictides with Mn$Pn$ ($Pn = \rm{P, As, Sb, Bi}$).  The Sb2 atoms bridge Mn atoms and provide the in-plane Mn-Mn superexchange paths (similar to As sites in BaMn$_2$As$_2$ and in other Mn$Pn$ systems \cite{Islam2020}).

Magnetization measurements were performed by using a Physical Property Measurement System (PPMS, Quantum Design) equipped with a vibrating sample magnetometer (VSM). For the temperature-dependent magnetization measurements, the samples were cooled down to the desired temperature with or without an applied magnetic field, termed as FC and ZFC, respectively. The temperature-dependent magnetization data were then collected upon warming at 2 K/min and at a fixed field. The magnetic field $H$ was applied either perpendicular to the MnSb layers (i.e., perpendicular to the plate; $H_\perp$) or  the $bc$ MnSb planes  ($H_{||}$). 

To determine the magnetic structures, single-crystal neutron diffraction experiments were carried out on the triple-axis spectrometer (TRIAX) located at the University of Missouri Research Reactor (MURR). The TRIAX measurements operate at an incident energy of $E_i = 14.7$ meV using a pyrolytic graphite (PG) monochromator system and is equipped with a PG analyzer stage. PG filters were placed before and after the second monochromator to reduce higher order contamination in the incident beam achieving a ratio  $I_\frac{\lambda}{2}:I_\lambda \approx 10^{-4}$. The beam divergence was defined by collimators of $60^\prime - 60^\prime- 40^\prime -80^\prime$ between the reactor source to a monochromator, monochromator to sample, sample to analyzer, and analyzer to detector, respectfully.   Crystals were  anchored to the cold tip of the Advanced Research Systems closed-cycle helium refrigerator and cooled to a base temperature of 6.7 K samples oriented in their (0$kl$) scattering plane.

\begin{table}
\caption{Lattice parameters $a$, $b$, $c$, the N\'eel temperature $T_{\rm N}$, and the average ordered magnetic moment $\langle gs \rangle$ of the \parent, \Zndop, \Cudop\ (the latter two are denoted as Zn-doped and Cu-doped, respectively) }
\renewcommand{\arraystretch}{1.0}
\setlength{\tabcolsep}{0.9em}

\begin{tabular}{llll}
\hline
\hline
         \tiny		& SrMnSb$_2^*$ 	& Zn-doped		& Cu-doped  \par	\\ \hline
$a$  ({\AA})     & 23.011(4)	& 23.144(5)	& 23.144(5) 	\\
$b$    ({\AA})   &  4.384(5) 	& 4.374(9) 	& 4.364(9)  	\\
$c$    ({\AA})   & 4.434(7) 		& 4.423 (9)   	& 4.440(9)  	\\
$T_{\rm N}$ (K)	& 300		& 205(5)		& 210(10)	\\
$\langle gs \rangle$    ($\mu_B$)   & 3.8(2)   	& 3.9(3)  		& 4.0(3)  	\\
\hline \hline
\label{tab:params}
\end{tabular}

* The values are compiled from Refs. \cite{Liu2017,Liu2019} 
\end{table}

To better understand and rationalize the experimental results of the SrMnSb$_2$ based materials at the atomic level, we have performed ab-initio density functional theory calculations based on the structure shown in Fig.\ \ref{Fig:structure}. To introduce chemical substitutions of the Mn site with either Zn or Cu, we considered supercells with various substitution levels, as discussed below.  In our DFT calculations, we have employed the generalized gradient approximation (GGA) for the exchange and correlation functional within the projector augmented wave (PAW) method \cite{Kresse1999,Kresse1996a} in conjunction with spin-orbit coupling (SOC), as implemented in the Vienna Ab-initio Simulation Package (VASP). We have used the experimental lattice parameters \cite{Blokhin2018} {see Table\ \ref{tab:params}  for the parent-SrMnSb$_2$ having space group of $Pnma$ and relaxed the structure for both atomic positions and lattice parameters. The theoretically optimized lattice parameters of parent-SrMnSb$_2$ are $a = 23.004$, $b= 4.4065$, and $c=4.4439$ {\AA}. To be as close as possible to the experimental 10\% substitution, we consider 12.5\% substitution of Cu and Zn, by exchanging one Mn atom in a $2\times1\times1$ supercell. For the supercell calculations we used the experimental lattice parameters. We also explore the higher 25\%  substitution by exchanging one of the four equivalent Mn sites in the chemical cell. The convergence criterion for the self-consistent calculations is $10^{-7}$ eV for the total energy per supercell, and an energy cutoff of 330 eV is used for the electronic wave functions. A $\Gamma$-centered grid of 4x12x12 k-points is used for Brillouin zone sampling in the self-consistent calculations.

\section{Results and Discussion}

Figure\ \ref{Fig:Mag} (a) shows the temperature dependence of the magnetic susceptibilities at an applied field of $H=1$ T  with $H$ perpendicular to the MnSb planes  ($\chi_{\perp}(T) \equiv M_{\perp}/H$) for \parent,   \Cudop\, and \Zndop. The step-like feature in $\chi_{\perp}(T)$ of the parent compound has been identified as a transition to the AFM structure depicted in Fig. \ref{Fig:structure}  at $T_{\rm N} = 300 \pm 3$ K, consistent with previous reports \cite{Liu2017,Liu2019}. The doped \Cudop\ and \Zndop\ show a weak feature at lower temperatures suggesting substantially lower $T_{\rm N} = 200 \pm 10$ and $210 \pm 12$ K, respectively. We note that spin-wave theory and the T-matrix approximation of a spin-$\frac{5}{2}$ quantum Heisenberg antiferromagnet on a 2D square lattice diluted with spinless vacancies predict $T_{\rm N}(x) \approx (1-2.6x)T_{\rm N}(0) = 220$ K ($x=0.1$ is the concentration of Cu/Zn) \cite{Chernyshev2002,Cowley1980}.  This is consistent with our observations of a reduced {\Tn}, including the determination from the neutron diffraction results.  Furthermore, the DFT, discussed below, show that the Cu or Zn reside on the Mn site and that both are spinless ($S\eqsim 0$).  This demonstrates that the substitution of Cu or Zn is random at the Mn site, i.e. there is no evidence of substitution of other atoms in the unit cell, and also that there is no apparent aggregation of the dopants. It is interesting to note that $\chi_{||}(T)$ (for $H$ in the $bc$-plane), shown in Fig. \ \ref{Fig:Mag} (b), is smooth at all temperatures with no obvious anomalies corresponding to $T_N$ such as those observed in $\chi_{\perp}(T)$.  This is typical of quasi two-dimensional (2D) AFM systems for which strong 2D correlations  develop at a much higher temperature than  that of the 3D transition temperature. Indeed, {in \parent}, 2D AFM correlations are observed up to 560K \cite{Liu2019b}.

Magnetization versus applied field data for $H_{\perp}$ ($M_\perp(H)$) (Fig.\ \ref{Fig:Mag} (c)) at $T= 2$ K   shows dHvA oscillations for \parent\  and \Zndop\ compounds but not for \Cudop.  By contrast, magnetization  curves versus applied field parallel to the planes $M_{||}(H)$  at $T= 2$ K  (Fig.\ \ref{Fig:Mag} (d))  for the three samples show linear dependence on $H$ as expected from slight moment canting for  typical AFM systems with moments that are oriented perpendicular to the applied field. This 2D behavior of the dHvA oscillations, observed only when the applied magnetic field is perpendicular to the Mn-layer, confirms that small Fermi surface pockets themselves are 2D in nature, forming cylinders along the $a$-axis.

Figure\ \ref{Fig:dHvA} (a) shows the magnetization of \Zndop\ (for $H_{\perp}$),  after removing a ferromagnetic impurity component (as described in more detail in Refs. \cite{Liu2019,Liu2019b}) versus the inverse of the applied field  $\Delta M(1/H)$ at various temperatures.   Direct numerical Fourier transforms (FT)  of $\Delta M(1/H)$ yield the FT spectra shown in Fig.\ \ \ref{Fig:dHvA} (b).  Figure\ \ \ref{Fig:dHvA} (c)  shows the temperature dependence of the integrated intensity over the main peaks ($H_{FT} \sim  74 $ T ) in the FT spectra (integrated from 40 to 120 T). The solid line is obtained by fitting the data to the Lifshitz-Kosevich equation to obtain the charge carrier's effective mass ($m^*$) as given in Ref. \cite{Kartsovnik2004}
\begin{equation}
P_{FT}= C\frac{K Tm^*/H_{\rm av}}{\sinh{(K Tm^*/H_{\rm av})}}
\end{equation}
where $C$ is a scale factor and $K = 2\pi^2k_B m_e/(\hbar e) \eqsim 14.69$ T/K. $H_{\rm av}$ corresponds to the inverse average of the field window over which the Fourier analysis is performed, such that $H_{\rm av}=[(1/ H_{\rm start}+1/ H_{\rm end})/2]^{-1} = 7.2$ T with $H_{\rm start}=6$ and $H_{\rm end}=9$ T. This is the same method used in Ref. \cite{Liu2017}. The fit to the data in Fig. \ref{Fig:dHvA} (c) yields a very small effective mass of $m^*=0.03(1)$, which is consistent with the results in Refs. \cite{Liu2017,Liu2019}, for \Ksub\ .

Figure\ \ref{Fig:dHvA} (d) shows the FT spectra of the dHvA data at $T = 2$ K for the \Zndop\ superimposed with those of \parent, and \Ksub\ (data adapted from Refs. \cite{Liu2019,Liu2019b}). We note two observations: first, the FT's center of gravity of \Zndop\ shifts to a higher field compared to that of the parent compound, and second, the FT is more structured and is likely a superposition of a few peaks. The more structured spectra of \Zndop\ suggest a more complicated electronic structure or may result from the inhomogeneinty of Zn distribution in MnSb layers.  

\begin{figure}
\includegraphics[width=3.3 in]{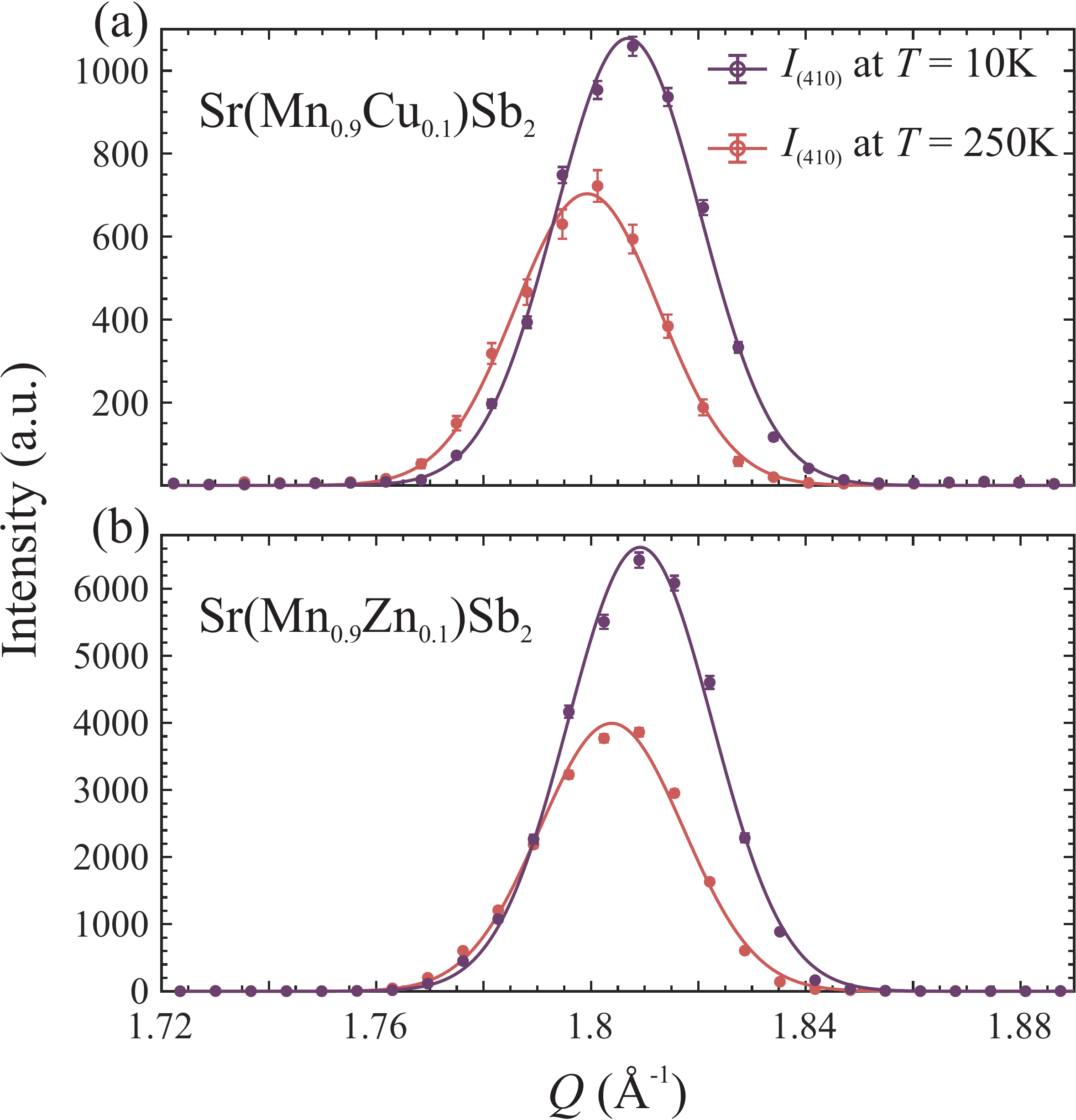}
\caption{Longitudinal scan at the (410) peak for (a) \Zndop\ and (b) \Cudop\ at  $T =10$ and 250 K showing temperature-dependent magnetic scattering, consistent with the C-type AFM structure of both systems.}
\label{Fig:Peaks1}
\end{figure}

\begin{figure}
\includegraphics[width=3.1 in]{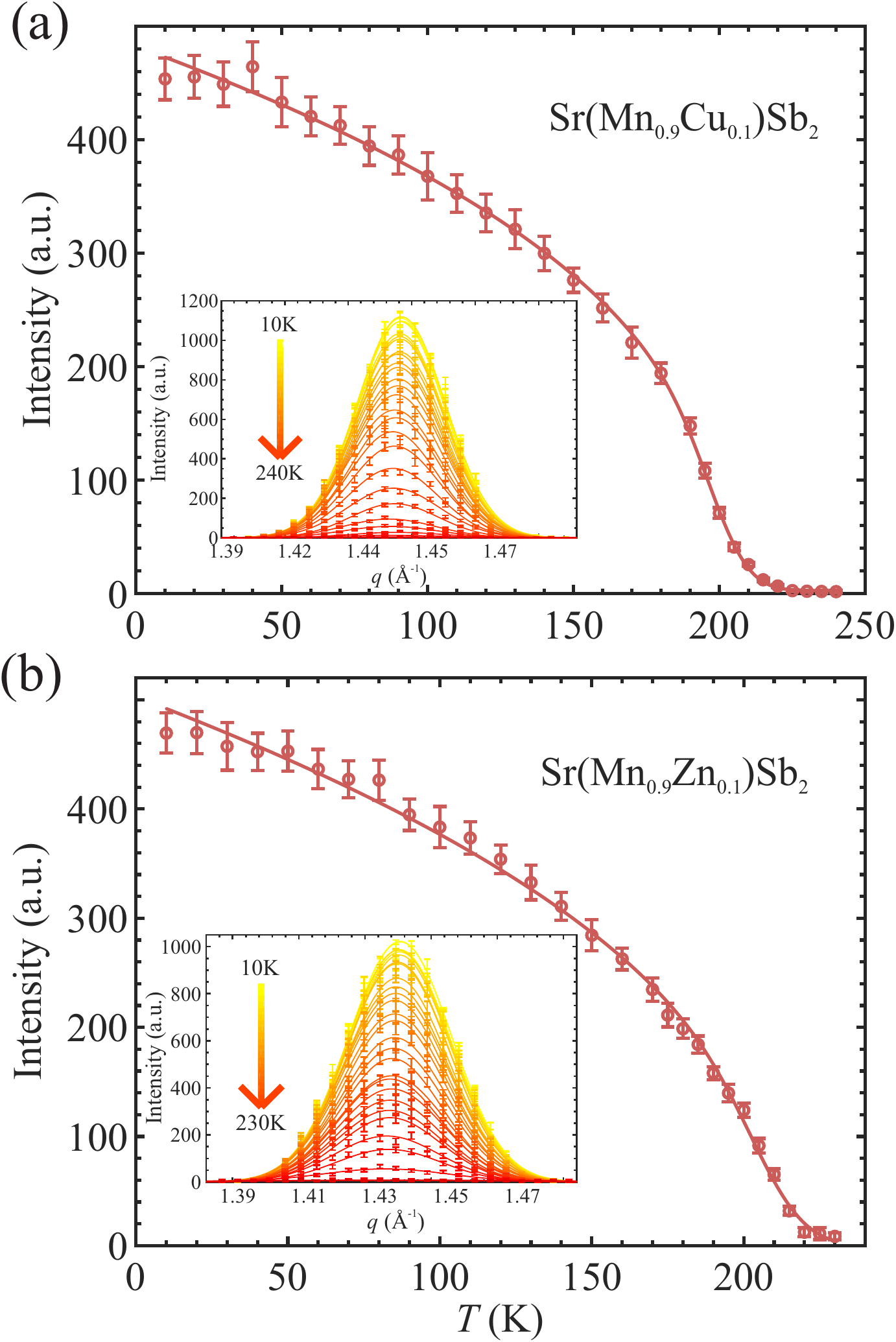}
\caption{Integrated intensity versus temperature of the magnetic (010) Bragg peak fitted to a power law $I\propto(1-T/T_N)^{2\beta}$ (solid line) for (a) \Cudop\  and (b) \Zndop\ with $\beta\approx0.20$ and $0.22$, and $T_N = 200\pm10$ and $210\pm12$ K, respectively. The inset shows the intensity of Bragg reflection (010) as a function of momentum transfer that demonstrates a decreasing trend in intensity as we increase temperature.}
\label{Fig:Peaks2}
\end{figure}

At base temperature ($T \sim 10$ K), the neutron diffraction data for both the \Zndop\ and \Cudop\ crystals show an intense (010) peak that is not allowed by the chemical symmetry of these crystals, indicating magnetic origin.  We also observe significant increase in the intensity of various low index nuclear (210), (410), and (610) Bragg peaks as the temperature is lowered (see Fig.\ \ref{Fig:Peaks1}).  The intensities at (010) and additional intensities on nuclear peaks disappear above the N{\'e}el temperature determined from susceptibility measurements, and are consistent with a simple C-type AFM structure similar to that of the parent compound \cite{Liu2017,Liu2019}. We note that ($h$00) peaks do not show appreciable temperature dependence, indicating limited or no canting of the moments away from the $a$-axis (we estimate an upper limit of a canted moment is less than $\sim 0.01$ and 0.02 $\mu_B$ per Mn site for both the Cu- and Zn-doped systems, respectively). Our detailed analysis of the magnetic peak intensities, assuming the moments are AFM in the plane and all stacked in the same way (C-type AFM),  allows us to determine the magnetic moment per Mn site as listed in Table\ \ref{tab:params}.    

Figure\ \ref{Fig:Peaks2}  shows the integrated intensities of the (010) magnetic Bragg reflections versus temperature with a fit (solid lines) to a power law $I \propto (1-T/T_N)^{2\beta}$   and assuming a gaussian distribution of $T_{\rm N}$.  We assume a distribution of $T_{\rm N}$ to account for possible inhomogeneities of Zn or Cu in the MnSb planes.  The fit to the data shown in Fig.\ \ref{Fig:Peaks2} yields  with $\beta\approx0.20$ and $0.22$, and $T_N = 200\pm10$ and $210\pm12$ K for the Zn-and Cu-doped crystals, respectively. The insets in Fig.\ \ref{Fig:Peaks2} show peak intensity versus longitudinal momentum transfer of the (010) magnetic peak at various temperatures. As described above, the suppression of $T_N$ relative to the parent \parent\ compound is consistent with site dilution at the level of $\approx$ 10\%.

\begin{figure}
\includegraphics[width=3 in]{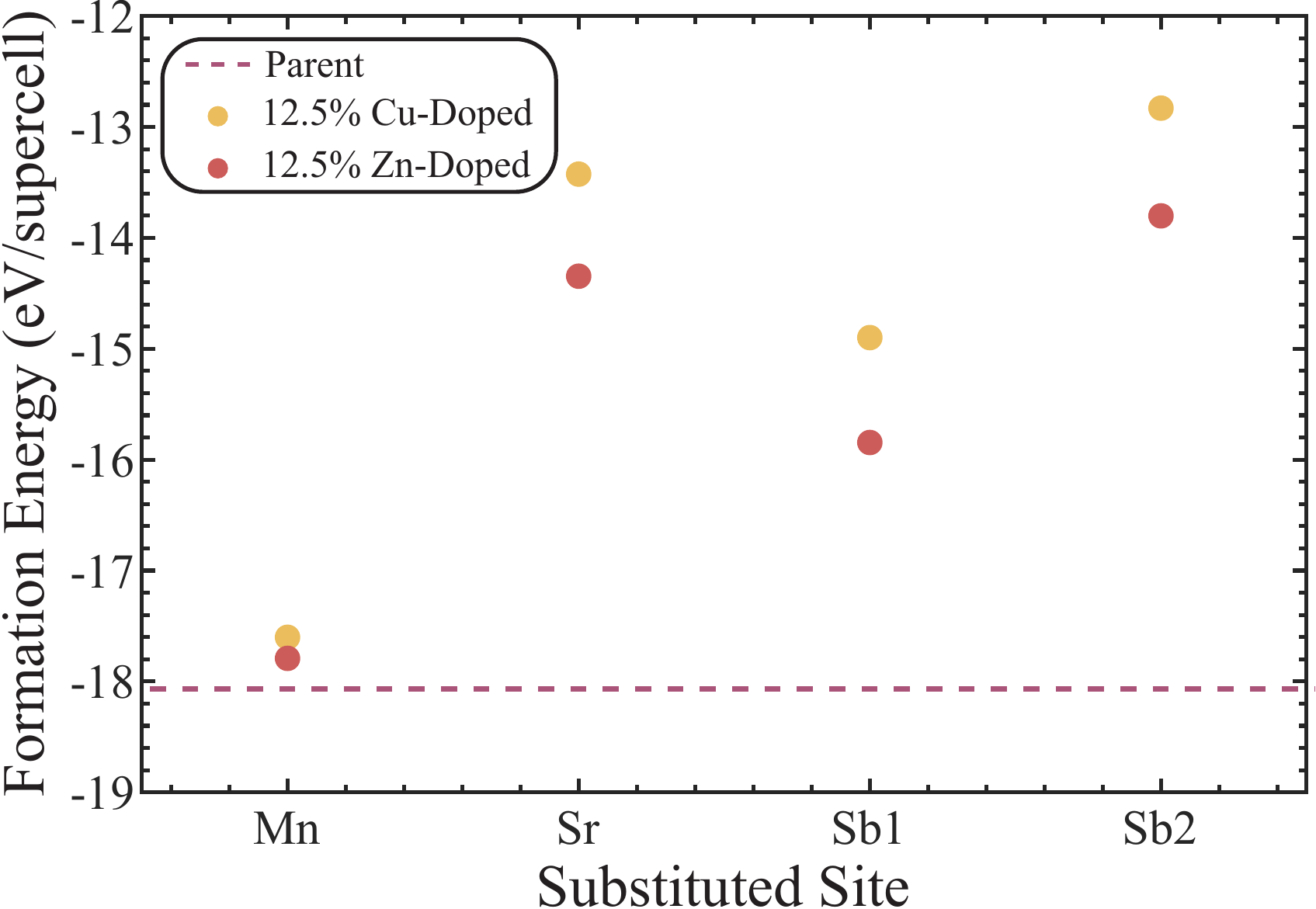}
\caption{Formation energy (FE) per supercell (i.e., $2\times 1\times 1$) for substitution of various sites (Mn, Sr, Sb1, or Sb2) of the parent \parent\ with a Cu or Zn atom. The FE of each substitution is compared to that of the parent compound (dashed line), which clearly shows that the Mn site is the most favorable in the formation of the doped system.}
\label{Fig:Total-E}
\end{figure}

\begin{figure*}[tp]
\includegraphics[width=6 in]{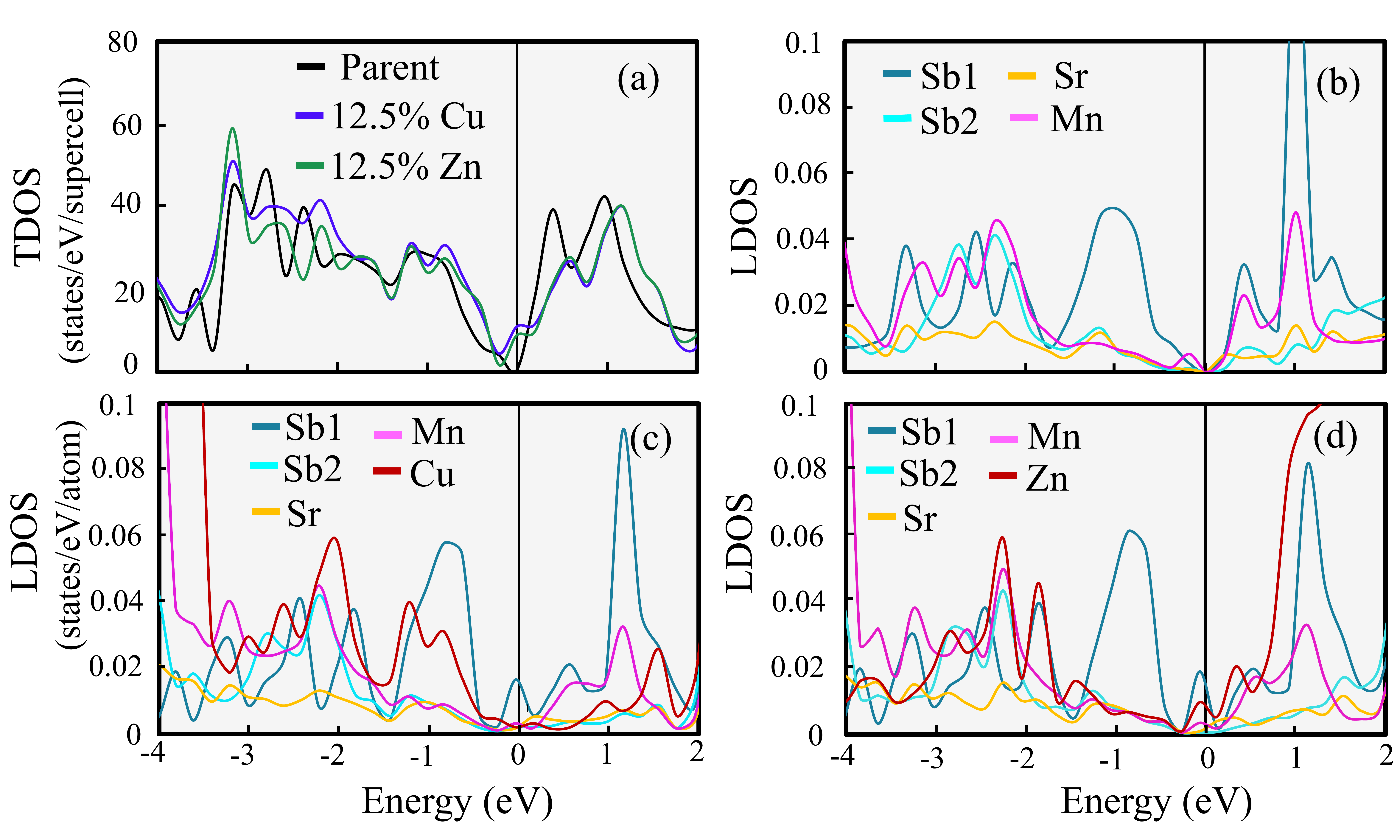}
\caption{(a) A superposition of the total density of states (DOS) of \parent\ and 12.5\% Cu/Zn-doped systems showing the parent's semi-metallic character and the shift of the Fermi level ($E_F=0$) by electron doping into the valence band, rendering the Cu/Zn doped systems metallic.  (b)  Atom specific partial DOS contributions to the total DOS for the parent compound.  Partial DOS curves with trends as a function of energy imply hybridization (for instance, Sb2-Mn) that have been implicated in invoking the in-plane Mn-Mn superexchange via $p$-states of Sb2 and $d$-states in Mn.  Note the substantial contribution of Sb1 to DOS near the Fermi level in the conduction and valence bands.  Energy bands associated with Sb1 have been implicated as the ones rendering the system ``topological.''  (c) Same as (b) for the $2\times 1 \times 1$ supercell with substitution of one Mn with Cu atom (12.5\% sample) showing the strong partial contribution to DOS by  Cu and Sb1 at and near the Fermi level. (d) Same as (c) for the Zn-doped sample with a strong contribution to the DOS from Sb1 and Mn.}
\label{Fig:DFT_DOS}
\end{figure*}

\section{DFT Calculations}
We employ DFT calculations to obtain a better understanding of the evolution of the magnetic properties of Cu and Zn doped \parent. First, we perform a detailed study of the parent-SrMnSb$_2$ to establish the band structure and magnetic structure. Various spin configurations of {\parent} - such as FM; C-, A-, and G-type AFM - are compared by calculating their ground state energies. C-type AFM configuration is energetically stable, but the energy difference between G-type and C-type AFM is very small $\sim 1.29$ meV/Mn, consistent with quasi 2D magnetism, as discussed previously \cite{Liu2019}. We also compare the ground state energy of the C-type structure with moments aligned either along the $a$, $b$, or $c$-axis. The energy difference between  Mn-moment along the $b$- and $c$-axis (within the MnSb-layer) is very small ($\approx0.03$ meV/Mn) with an insignificantly  lower value for the moment along the $b$-axis. The energy is the lowest for moment orientation perpendicular to the Mn layer (i.e., along the $a$-axis), consistent with the experimentally determined C-type AFM structure (See Table \ref{tab:DFTparams} for more details).

We employ similar calculations to the 12.5\% Cu- and Zn-doped systems with $2\times 1\times 1$-supercell, as shown in Fig. \ref{Fig:structure}. As discussed above regarding the reduced \Tn, it is justified to assume that Cu and Zn substitute the Mn site only. To confirm that, we conducted formation energy calculations of  a supercell with  Cu or Zn in the Mn, Sr, Sb1, and Sb2 sites and found that the lowest energy is achieved  in the substitutions of the Mn site (see Fig. \ref{Fig:Total-E} showing Mn site substitution has the lowest formation energy). The formation energies are calculated by subtracting total energies of the elements with the appropriate content from the total energy of the doped compound. This further justifies our ensuing  detailed calculations by placing the Cu or Zn substitutions only on the Mn site.

We find that the C-type AFM structure is also the most stable configuration in the substituted samples. Based on the observations of the parent compound,  we compared the ground state energies of doped systems in the C-type AFM structure with Mn-moment along the $a$- and $b$-axis only.  Table\ \ref{tab:DFTparams} lists the ground-state energy differences for the two spin directions. From the energy differences between Mn-moment aligned along the $a$-axis and $b$-axis ($\Delta E_{ab}$) of all the systems, we find that Mn-moment is aligned along the $a$-axis for 12.5\% Zn/Cu-doped systems.  Similar considerations  for the higher concentration of 25\% (chemical unit cell with substitution of one Mn with Cu or Zn) show that the C-type AFM is preserved with Mn-moment  along the $a$-axis for the Zn-doped, whereas for the Cu substitution, it is reoriented along the $b$-axis. This is expected as Mn$^{2+}$, with an effective total orbital moment $L \eqsim 0 $, exhibits relatively weak single-ion anisotropy, and thus is susceptible to spin-flop either by external magnetic field or by local atomic environment.

\begin{figure*}[tp]
\includegraphics[width=6 in]{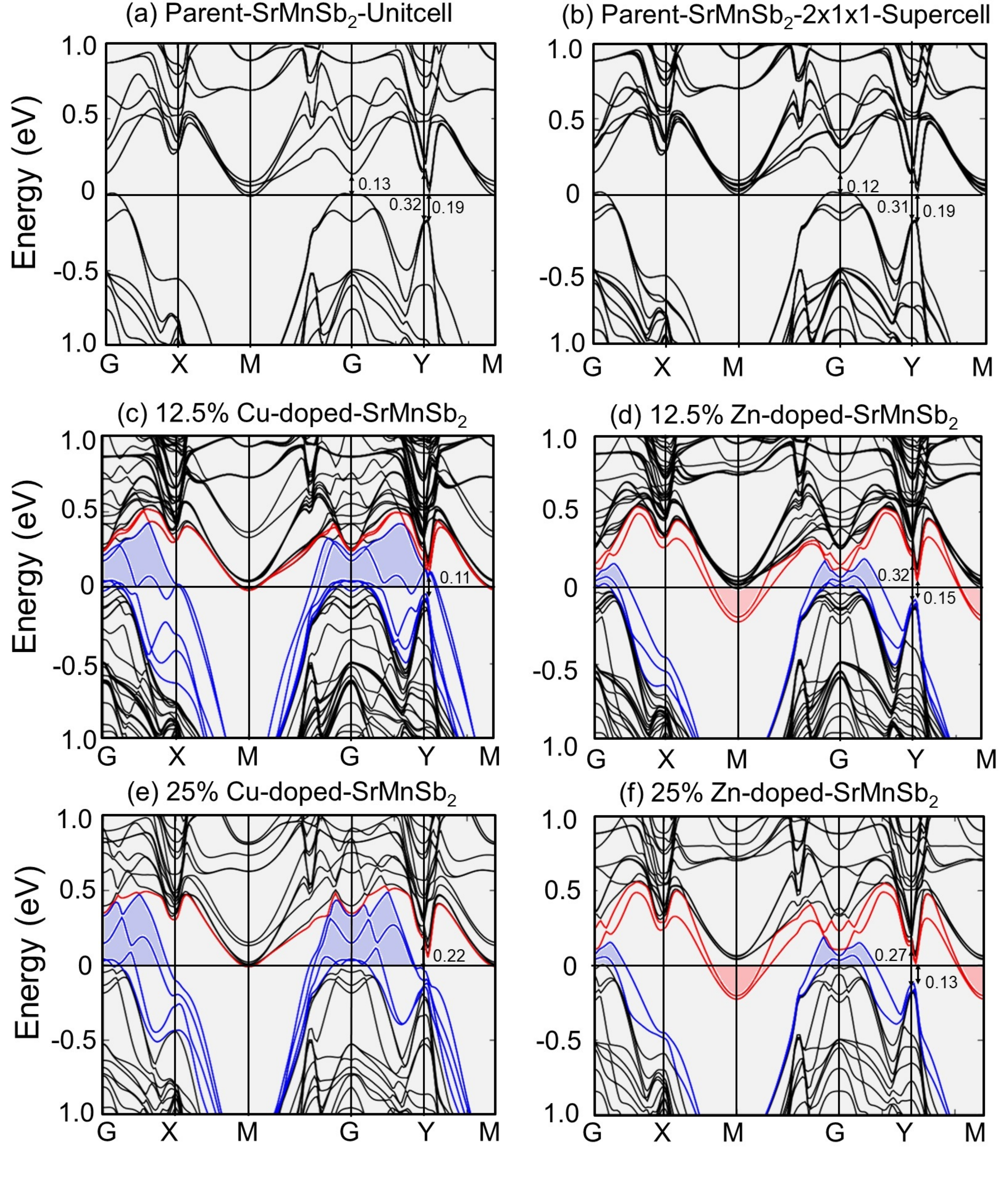}
\caption{Band structures of (a) a chemical unit cell and (b) a $2\times1\times1$-supercell of parent SrMnSb$_2$, and (c, e) doped systems  with 12.5\%, 25\% Cu, and (d, f) 12.5\%, 25\% Zn. The shaded light blue and light red colors represent the regions that are shifted upon doping from the VB to CB (hole-pockets) and from CB to VB (electron pockets), respectively. The Fermi level is at $E = 0 $ eV. The gap of the Dirac cone and the gap at the Y-point is shown using an arrow. Red and blue bands around the Fermi level are the crossed bands from CB to VB and vice-versa, respectively, due to the doping effect.}
\label{Fig:DFT_Band}
\end{figure*}

 \begin{table}
\caption{The energy difference between C-Type parent,  Cu-, and Zn-doped systems with Mn-moment aligned along the $a$-axis and $b$-axis ($\Delta E_{ab} \equiv E_{b{\rm-axis}}-E_{a{\rm -axis}}$), and magnetic moment of Mn. Note that for the 25\% Cu-doped system, the magnetic moment flops into the Mn-plane.}
\renewcommand{\arraystretch}{1.0}
\setlength{\tabcolsep}{0.9em}
\centering

\begin{tabular}{lll}
\hline
\hline
Systems				  	& $\Delta E_{ab}$		& Mn-moment \\
						& (meV/Mn)						& ($\mu_B$/Mn) \\ \hline
Parent		     		& 0.23			& 3.75		 	\\ \hline
12.5\% Cu-Doped		   	& 0.15			& 3.84  		\\
25\% Cu-Doped			& -0.75			& 3.73			\\ \hline
12.5\% Zn-Doped			& 0.27 			& 3.83		  	\\
25\% Zn-Doped			& 1.6			& 3.78			\\

\hline \hline
\label{tab:DFTparams}
\end{tabular}
\end{table}

The calculated average Mn-moment in the parent, 12.5\% Cu-doped, and 12.5\% Zn-doped systems is 3.75, 3.84, and 3.83 $\mu_B$ per Mn atom, respectively,  agreeing with the presented experimental values (see Table\ \ref{tab:params}). Our calculations also determine that the average magnetic moments on the Cu and Zn sites  are both negligible  with 0.02 $\mu_B$ per Cu and  and 0.01 $\mu_B$ per Zn.  These results imply that effectively Cu$^{+}$ substitutes the Mn$^{2+}$ site, and Zn$^0$ or Zn$^{2+}$ substitutes the Mn$^{2+}$ site.

\begin{figure*}[]
\includegraphics[width=6 in]{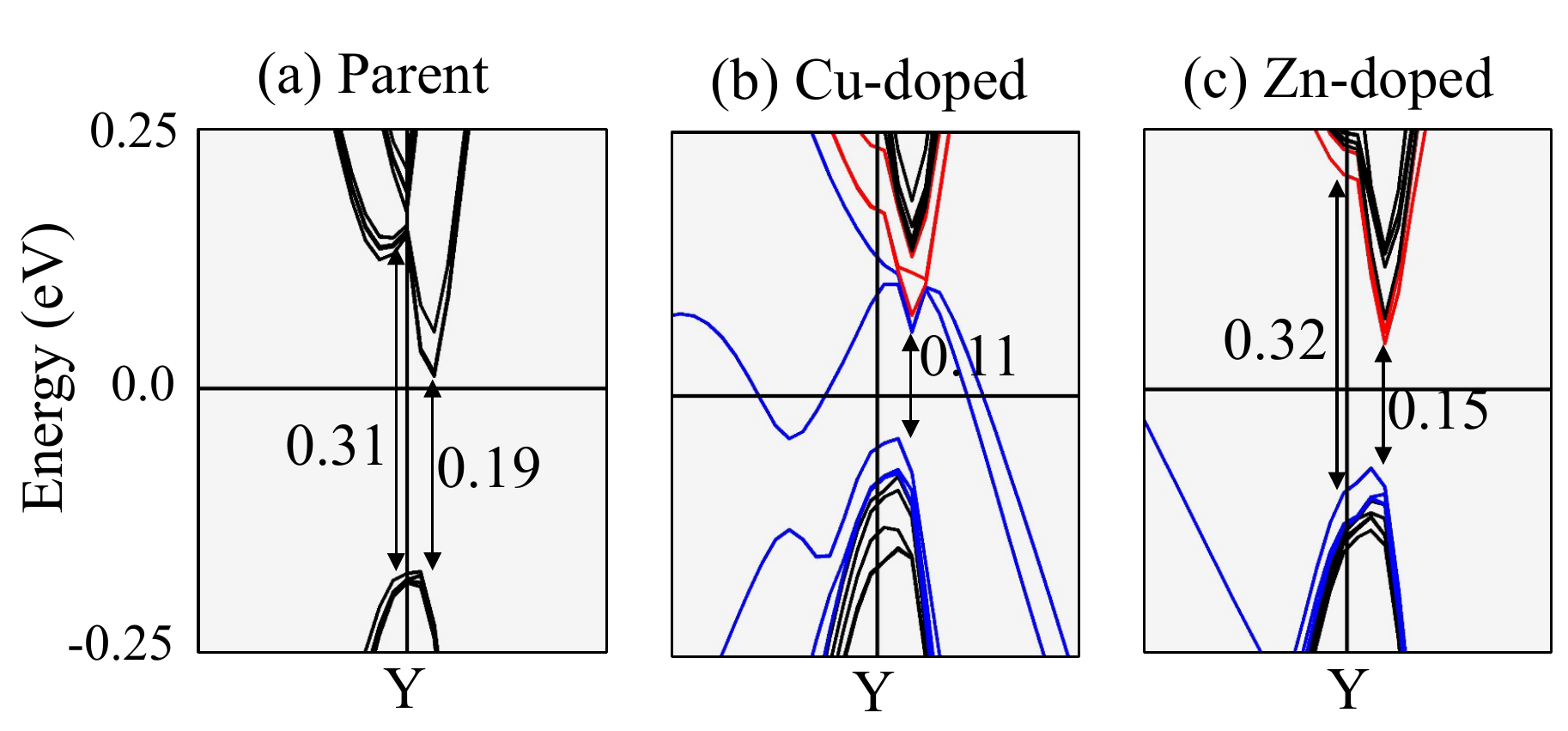}
\caption{Band structures calculations focused on the Y-point near the Fermi level for  (a) parent, (b) 12.5\% Cu-doped, and (c) 12.5\% Zn-doped systems. The gaps at the Dirac cone and the Y-point are indicated by arrows, and the values are in eV. The Dirac feature near Y-point and the corresponding gap in the Zn-doped system, shown in (c), are almost as those of \parent, as shown in (a). The  Cu-doped system, shown in (b), has significantly distorted bands at the Y-point, making the system less Dirac-like, and likely inhibiting the dHvA effect.  Blue solid lines in (b) and (c) show band-crossing at the Fermi level from VB to CB. }
\label{Fig:Y-point}
\end{figure*}

Figure \ref{Fig:DFT_DOS} shows the density of states (DOS) of the parent and 12.5\% Zn/Cu doped systems.  The  Fermi level of the parent compound  resides in a narrow minimum of the DOS consistent with the semi-metallic (or narrow semiconducting) character of \parent.  Upon Cu/Zn substitution, the Fermi level shifts into the valence band (VB). The Cu-doped system has a slightly higher DOS  at the Fermi level compared to the Zn-doped system.

Atom projected DOS (Fig. \ref{Fig:DFT_DOS} (b-d)) of each system is also compared  to ascertain the major contribution of each element at and near $E_F$.  It is evident that for \parent, Sb1 $p$-states are the main contributor to the DOS near the conduction and valence bands.  Figure \ref{Fig:DFT_DOS} (b) shows hybridization of Sb1 $p$-states with Mn $d$-states above the Fermi level ($\sim 0$ to 1 eV range), whereas Mn hybridizes strongly with Sb2 below the Fermi level ($\sim  -2$ to $-4$ eV range).   It is interesting  to note some  hybridization of Cu  and Sb1 states in this energy range, which is absent in the Zn doped system. This indicates different effects of Cu- and Zn- impurities on the electronic structure pointing out the possibility that a Cu site introduces a larger cross-section than Zn site for scattering free carriers in the Sb1 layer.  Such scattering may affect the lifetime of a free carrier (i.e., shorten the mean free path), which is detrimental to the dHvA effect, and can rationalize the absence of the dHvA effect for Cu doped compounds.


Figure \ref{Fig:DFT_Band} shows the band structures of the parent, Cu-doped, and Zn-doped systems at concentration 12.5\% (replacing one Mn with Cu/Zn  in  $2\times1\times1$-supercell), and at 25\% (replacing one Mn with Cu/Zn in the chemical unit cell). Figures\ \ref{Fig:DFT_Band} (a) and (b) show band structure calculations of the \parent\ using the chemical unit cell and $2\times 1\times 1$-supercell, respectively  to ensure invariance of results to the choice of cell size.  All features in the band structure of $2\times1\times1$-supercell (Fig. \ref{Fig:DFT_Band} (b)), including the high symmetry points, are the same as those calculated for the chemical unit cell bands (Fig.\ \ref{Fig:DFT_Band} (a)), and agrees well with previous reports \citep{Ramankutty2018,Weber2018}. 
In the band structure of SrMnSb$_2$ (Fig. \ref{Fig:DFT_Band} (a-b)), the Fermi level touches the VB at the G-point, and the CB at the M-point.   There is a Dirac cone near the Y-point, albeit with a band-gap of $\sim 0.19$ eV due to spin-orbit coupling, and a 0.31 eV gap at the Y-point (for more detailed band structure around the Y-point, see Fig. \ref{Fig:Y-point} (a)). 
The calculation here suggests that the Fermi surfaces of \parent\ are hole-like (or have hole-pockets) at the G-point and electron like (or electron pockets) at the Y- and M-points \cite{Singleton2001}. However, recent Hall coefficient measurement, ARPES, and quantum oscillations of \parent\ are consistent with a lower Fermi level and a hole-like pocket at the Y-point \cite{Liu2019,Ramankutty2018}.

Figures \ref{Fig:DFT_Band} (c) and (d) show the band structures of the 12.5\% of Cu/Zn substituted in a $2\times1\times1$-supercell, showing that Cu and Zn substitutions change the features at the high symmetry points differently. The doped systems are metallic-like in nature, as also demonstrated above from the DOS. Cu substitution mostly affects the bands at G, Y, and X-points, whereas Zn affects only  the M and G-points. Most interesting is the fact that the Dirac cone near the Y-point is practically not affected by Zn-doping, whereas it is a significantly changed with Cu-doping, as can be clearly seen in Fig.\ \ref{Fig:Y-point}. Figure \ref{Fig:DFT_Band}  (d) shows that the 25\% Cu/Zn-doping has almost the same effects on the energy bands as the 12.5\% doping. Cu substitution affects the Y-point significantly enough to distort the Dirac cone and diminish the dHvA oscillations, as observed experimentally.}    

\section{Conclusions}
We have grown single crystals of {\Cudop} and  {\Zndop},  and determined their properties using magnetic susceptibility, magnetization versus the applied magnetic field, and neutron diffraction measurements. We find that the Mn$^{2+}$ moments in both systems undergo  C-type antiferromagnetic ordering at $T_N = 200\pm10$ and $210\pm12$ K, respectively - which is significantly reduced compared to their parent  SrMnSb$_2$ with $T_{\rm N}=297 \pm 3$ K. The reduction of $T_N$ with respect to the parent compound is likely caused by dilution effects. Analyzing the neutron diffraction patterns, we find that the magnetic moment per Mn$^{2+}$ is 3.9(3) and 4.0(3) $\mu_B$ for the Cu- and Zn-doped systems and slightly larger than that of the parent \parent. 
Magnetization versus applied magnetic field (perpendicular to the MnSb planes) measurements at low temperatures (below $\sim 30$ K) show oscillations due to the dHvA effect for the Zn-doped system, with features that are slightly modified compared to those of the parent compound. In contrast, the Cu-doped system does not show the dHvA effect.  
 
We also provide DFT calculations including spin-orbit coupling which confirm that Zn and Cu substitute for Mn. DFT predicts C-type AFM order with a magnetic moment that agrees with experiments in the parent, and doped Cu and Zn systems. The calculations also identify the easy $a$-axis of the magnetic moment in the parent and 12.5\% of Cu or Zn compositions. Interestingly, for the 25\%  Cu content, the calculations predict an easy  $b$-axis C-type antiferromagnet, which has not yet been observed experimentally. Considering that the single-ion anisotropy is expected to be very weak for an Mn$^{2+}$ ($d^5$ configuration) with $L \eqsim 0$, it is likely that the spin direction can be readily flopped by either magnetic field or internal perturbation (i.e., crystal field effects), in this case, apparently exerted by the Cu substitution.  

The DFT calculations show that the incorporation of Cu and Zn in SrMnSb$_2$ causes distortions of the electronic bands at the Fermi level. The calculations yield two distinct behaviors of the Cu and the Zn doped systems that qualitatively rationalize the absence and observation of the dHvA effect, respectively. First, the calculated DOS show that the Cu states are more hybridized with Sb1's $p$-states and are more abundant near the Fermi level, possibly increasing the carrier scattering, which results in a decrease in the mean free path of the carriers that will suppress the dHvA oscillations. Second, the band structure calculations for the Zn-doped system show that the Y-point, hosting a gapped Dirac cone and being responsible for the dHvA oscillations, is only slightly modified compared to the parent {\parent}. By contrast, the calculations for the Cu-doped system show severe disruption near the Y-point, which may inhibit the occurrence of the dHvA effect.

\acknowledgments

This research was supported by the U.S. Department of Energy, Office of Basic Energy Sciences, Division of Materials Sciences and Engineering.  Ames Laboratory is operated for the U.S. Department of Energy by Iowa State University under Contract No.~DE-AC02-07CH11358. A portion of this research used resources at the High Flux Isotope Reactor, a DOE Office of Science User Facility operated by the Oak Ridge National Laboratory.

The theoretical capabilities, specific to electronic (band) structure, magnetic order, and magnetic anisotropy predictions, employed in this work have been developed in Critical Materials Institute, an Energy Innovation Hub led by Ames Laboratory and funded by the U. S. Department of Energy, Office of Energy Efficiency and Renewable Energy, Advanced Manufacturing Office. R. C. and D. P. would like to acknowledge Ed Moxley for maintaining computational facilities, including the RAMAN cluster and computational software.

\bibliographystyle{apsrev4-2}
\bibliography{SrMnSb2}

\end{document}